\documentclass[acmsmall,screen]{acmart}

\usepackage{amsmath}

\acmJournal{TOMM}
\acmYear{2026}
\copyrightyear{2026}
\acmDOI{}
\setcopyright{none}

\begin{document}

\title[Packet-Loss Robust 3D Gaussian Compression]{Packet-Loss Robust 3D Gaussian Compression via Atomic Packaging and GNN-based Error Concealment}

\author{Yuxuan Tao}
\email{taoyuxuan2002@qq.com}
\affiliation{%
  \institution{Central South University}
  \city{Changsha}
  \country{China}
}

\author{Xuerui Ma}
\email{maxuerui@mlslabs.com.cn}
\affiliation{%
  \institution{Malanshan Audio \& Video Laboratory}
  \city{Changsha}
  \country{China}
}

\author{Hao Zhang}
\authornote{Corresponding author.}
\email{hao@csu.edu.cn}
\affiliation{%
  \institution{Central South University}
  \city{Changsha}
  \country{China}
}

\author{Chunhua Peng}
\authornote{Corresponding author.}
\email{pch@csu.edu.cn}
\affiliation{%
  \institution{Central South University}
  \city{Changsha}
  \country{China}
}

\renewcommand{\shortauthors}{Tao et al.}

\begin{abstract}
While 3D Gaussian Splatting (3DGS) and its high-efficiency compression schemes (e.g., HAC++) deliver state-of-the-art rendering performance under ideal network conditions, they remain highly vulnerable to packet loss during streaming. Existing compression methods encode scene attributes (features, scales, offsets) into independent bitstreams; consequently, the loss of a single packet results in structurally inconsistent, ``broken'' anchors that cause severe visual artifacts. To address this, we propose a robust 3DGS transmission and error concealment framework tailored for lossy networks. At the encoder, we introduce \textit{Anchor-level Atomic Packaging} to ensure spatial attributes are jointly encapsulated, guaranteeing that losses manifest solely as cleanly missing anchors rather than corrupted parameters. This is complemented by \textit{Stratified Random Grouping}, which disperses potential packet losses uniformly across the spatial domain to prevent large contiguous voids. At the decoder, we formulate error concealment as a prior-aware dual-branch inpainting task. We first establish a geometrically robust baseline via Context-Aware Residual Interpolation (CARI), and subsequently employ a lightweight two-layer Graph Neural Network (GNN), conditioned on hash-grid priors and cross-attention fusion, to regress high-frequency attribute residuals. An attribute-wise confidence module controls the GNN output and falls back to interpolation when the corresponding prediction is unreliable. Comprehensive experiments under 20\% random packet loss demonstrate that our framework maintains high perceptual quality and stability, achieving an overall average PSNR degradation of approximately 3 dB across the evaluated datasets, with larger drops on more challenging large-scale scenes. The GNN branch provides complementary gains in complex high-frequency scenes, while the confidence-controlled fallback preserves the stronger CARI estimate in simple or low-confidence regions. This work provides a foundation for robust 3DGS streaming, leaving joint optimization of bitrate and runtime efficiency for future investigation.
\end{abstract}



\maketitle

\section{Introduction}
\label{sec:introduction}

In recent years, 3D Gaussian Splatting (3DGS) has achieved breakthrough progress in 3D scene reconstruction and novel view synthesis, thanks to its high-fidelity and real-time rendering capabilities. However, the massive number of parameters in vanilla 3DGS---comprising millions of Gaussian primitives with attributes such as color, scale, and rotation---poses significant challenges for practical storage and network transmission \cite{chen2025hac}. Consequently, numerous compression algorithms for 3DGS have emerged. Notably, Scaffold-GS \cite{lu2024scaffold} introduced a sparse representation based on anchors, and the recent HAC++ \cite{chen2025hac} further achieved an extreme compression ratio of over $100\times$ through Hash-grid Assisted Context and decoupled encoding of anchor attributes, greatly advancing the practical deployment of 3DGS.

\begin{figure*}[htbp]
    \centering
    \includegraphics[width=\textwidth]{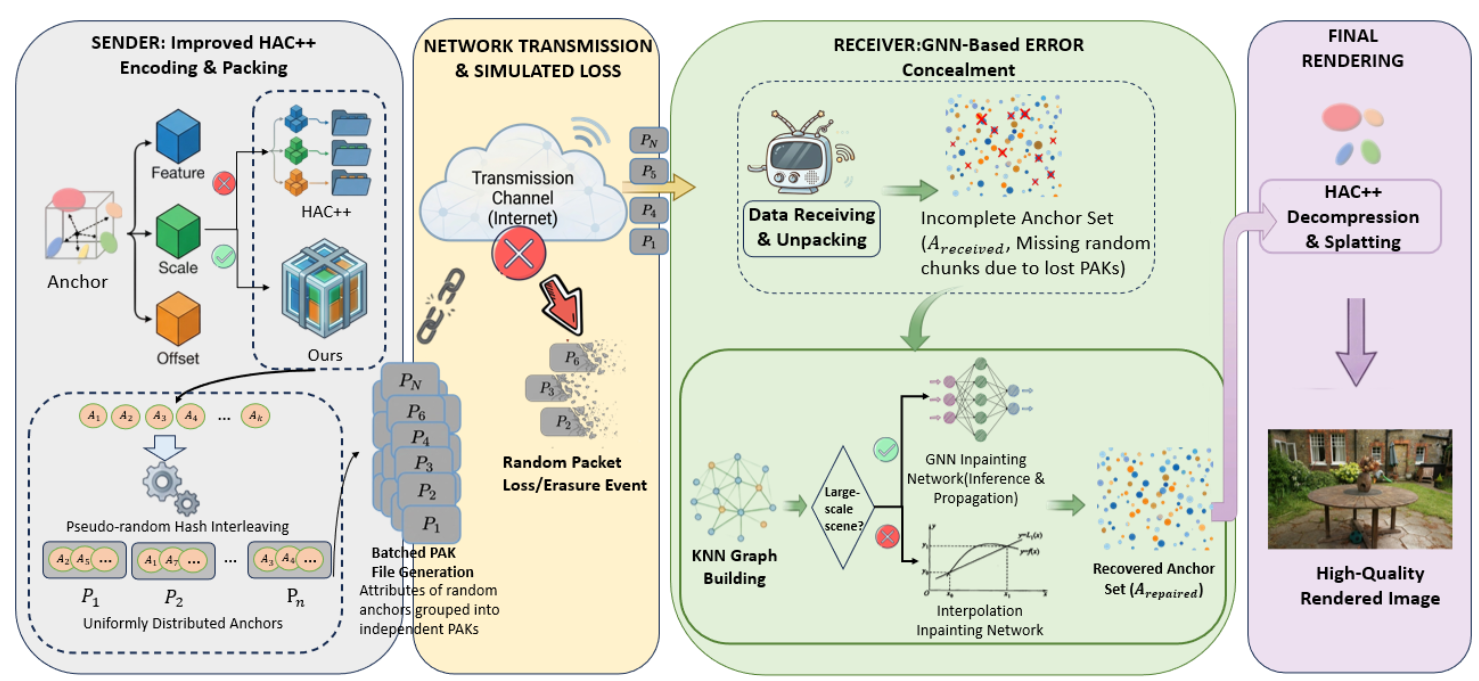} 
    \Description{Pipeline overview comparing attribute-decoupled packaging and Z-order grouping with the proposed atomic packaging and stratified random grouping, followed by CARI and GNN-based decoder-side recovery.}
    \caption{Overview of our proposed robust 3DGS transmission and error concealment framework. (Left: Encoder Strategies) Compared to the conventional failure mode (attribute-decoupled packaging and Z-order grouping) which causes catastrophic broken anchors and large spatial voids under packet loss, our encoder introduces \textbf{Atomic Packaging} and \textbf{Stratified Random Grouping} to ensure that losses strictly manifest as uniformly dispersed sparse missing anchors. (Right: Decoder \& Inpainter) To recover the missing queries, we construct a prior-aware dual-branch inpainting pipeline. The upper branch establishes a robust low-frequency baseline via \textbf{Context-Aware Residual Interpolation (CARI)}, while the lower branch employs a lightweight two-layer \textbf{Graph Neural Network (GNN)} to regress high-frequency details. Finally, \textbf{attribute-wise confidence control} blends or falls back to interpolation when a predicted attribute is unreliable, improving rendering stability under severe loss.}
    \label{fig:framework} %
\end{figure*}

Nevertheless, while highly efficient compression schemes like HAC++ excel under ideal channel conditions and offline storage, their rendering quality is severely compromised by the inevitable random and burst packet losses in real-world network streaming. To approach the Shannon entropy limit, existing HAC++ packaging strategies deeply decouple strongly correlated anchor attributes (features, scales, and offsets) into independent bitstreams \cite{chen2025hac}, while organizing anchors sequentially based on spatial proximity (e.g., Morton coding). Under lossy network conditions, this extreme compression design exposes two fatal vulnerabilities: First, the loss of any single attribute bitstream generates parameter-mismatched ``inconsistent broken anchors'' at the decoder. Due to the accumulative effect of alpha blending during rendering, these broken anchors induce visual artifacts far more destructive than the ``complete absence of anchors,'' as visually contrasted in the left panel of Fig.~\ref{fig:framework}. Second, when encountering burst packet loss, the sequential spatial organization creates massive, continuous spatial voids, rendering the decoder unable to leverage local spatial redundancy for interpolation-based recovery.

To address these fundamental vulnerabilities, we propose a robust 3DGS transmission and error concealment framework. Our encoder design is guided by a key insight: under alpha blending, retaining an anchor with corrupted attributes is strictly more destructive than its complete absence. We therefore introduce \textit{Anchor-level Atomic Packaging}, which jointly encapsulates all attributes of each anchor into the same transmission packet, mechanically guaranteeing an all-or-nothing failure mode. This is complemented by \textit{Stratified Random Grouping}, which distributes anchors across packets via hash-based interleaving, transforming potential burst losses into spatially uniform sparse absences rather than concentrated voids.

At the decoder side, we formulate error concealment as a prior-aware dual-branch inpainting task. The first branch, Context-Aware Residual Interpolation (CARI), decomposes known anchor attributes into hash-grid prior predictions and high-frequency residuals, then aggregates neighbor residuals through a multi-factor spatial--semantic weighting scheme to establish a geometrically robust baseline. The second branch employs a lightweight two-layer Graph Neural Network (GNN) \cite{wang2019dynamic}, conditioned on hash-grid priors via cross-attention fusion, to predict attribute residuals that capture fine-grained details beyond the reach of local interpolation. An attribute-wise confidence module controls the GNN output and falls back to interpolation when the corresponding attribute confidence is insufficient. This design is intentionally conservative: for smooth or small-scale scenes where interpolation is already close to optimal, the decoder can retain the CARI estimate instead of forcing a learned correction. An overview of the complete pipeline is depicted in Fig.~\ref{fig:framework}.

The main contributions of this paper are summarized as follows:
\begin{itemize}
    \item We reveal that, under the alpha-blending rendering pipeline, attribute-inconsistent ``broken anchors'' are fundamentally more destructive than cleanly missing anchors, and accordingly propose an anchor-level atomic packaging mechanism that physically guarantees an all-or-nothing failure mode, converting the intractable artifact removal problem into a well-defined point cloud completion task.
    \item We design a hash-based stratified random grouping strategy that distributes anchors uniformly across transmission packets, ensuring that burst packet loss manifests as spatially uniform sparse downsampling rather than catastrophic contiguous voids.
    \item We formulate decoder-side error concealment as a prior-aware dual-branch inpainting problem, integrating Context-Aware Residual Interpolation (CARI) with a lightweight two-layer GNN augmented by cross-attention over hash-grid priors, and introduce attribute-wise confidence control with fallback to interpolation when predictions are unreliable.
    \item We establish a packet-level 20\% random-loss evaluation protocol and demonstrate across three standard benchmarks that our framework reduces the overall average PSNR degradation to approximately 3~dB, substantially outperforming no-concealment transmission while using confidence control to combine GNN refinement with interpolation fallback.
\end{itemize}

\noindent \textbf{Organization:} The remainder of this paper is organized as follows: Section~\ref{sec:related-work} reviews related work. Section~\ref{sec:method} details the encoder-side strategies and the decoder-side error concealment pipeline. Section~\ref{sec:Experiments} presents the experimental setup, quantitative results, and ablation studies. Finally, Section~\ref{sec:conclusion} concludes the paper and outlines future directions.

\section{Related Work}
\label{sec:related-work}

\subsection{3D Gaussian Splatting and Compression}
Recently, 3D Gaussian Splatting (3DGS) \cite{kerbl20233d} has gained significant attention for its high-fidelity and high-frame-rate rendering capabilities. However, the massive parameter count poses a substantial challenge for practical deployment, especially for free-viewpoint and large-scale neural rendering scenarios previously studied in image-based rendering and progressive NeRF systems \cite{hedman2018deep,xiangli2022bungeenerf}. Consequently, numerous compression algorithms for 3DGS have emerged. In early explorations focusing on ``attribute value'' compression, Fan et al. proposed \textbf{LightGaussian} \cite{fan2024lightgaussian}, which draws inspiration from network pruning to eliminate redundant points by calculating the global importance of Gaussians to scene reconstruction, while utilizing knowledge distillation to transfer high-order Spherical Harmonic (SH) coefficients into a compact format. Subsequently, Niedermayr et al. introduced \textbf{Compressed 3D Gaussian Splatting} \cite{niedermayr2024compressed}, which incorporates sensitivity-aware vector clustering and quantization-aware training (QAT) to perform codebook compression on Gaussian color and geometric attributes.

However, these methods typically treat Gaussian points as independent primitives, overlooking the strong correlation between spatially adjacent points. To overcome this bottleneck, \textbf{Scaffold-GS} \cite{lu2024scaffold} first introduced an ``Anchor'' mechanism, utilizing local MLPs to dynamically predict Gaussian attributes. Building on this, \textbf{HAC} \cite{chen2024hac} and its successor \textbf{HAC++} introduced Hash-grid Assisted Context to deeply decouple attributes such as anchor features, scales, and offsets, implementing high-precision quantization and entropy coding independently to achieve extreme compression ratios. Nevertheless, such attribute-independent packaging strategies---designed for ``limit Shannon entropy''---are extremely fragile during real-world network transmission. The loss of any single attribute results in ``fragmented anchors'' with inconsistent parameters, leading to catastrophic visual artifacts.

\subsection{Error Concealment in Video Coding}
In traditional 2D video streaming, packet loss is an unavoidable reality. For intra-frame data loss, conventional spatial error concealment methods primarily leverage the spatial smoothness of boundary pixels for local interpolation or texture synthesis \cite{wang2000error}. 

To fundamentally prevent large-scale spatial holes at the decoder side due to packet loss, literature has long established the effectiveness of proactively reshaping data distribution at the encoder side. For instance, the \textbf{Flexible Macroblock Ordering (FMO)} mechanism in traditional video coding standards (e.g., H.264/AVC) \cite{wenger2003h} intentionally interleaves and scatters macroblocks---similar to a checkerboard---across different network packets. This ensures that a lost block is always surrounded by successfully received reference pixels. Furthermore, \textbf{Multiple Description Coding (MDC)} \cite{goyal2001multiple} distributes images across multiple description streams via spatial downsampling. The ``atomic packaging'' and ``stratified random grouping'' strategies proposed in this paper represent an innovative mapping of these two classic streaming fault-tolerance philosophies onto unordered 3DGS neural rendering primitives.

\subsection{Deep Learning for Point Cloud Completion}
For 3D point clouds that lack a regular 2D grid topology, traditional spatial interpolation methods often struggle to effectively recover missing high-frequency geometric details. Recently, utilizing neural networks for 3D point cloud processing and restoration has become a prominent research trend. Although \textbf{PU-GCN} \cite{qian2021pu}, proposed by Qian et al., targets point cloud upsampling, its core concept---leveraging Graph Convolutional Networks (GNNs) to aggregate local neighborhood features for generating new point coordinates and attributes---has proven effective for repairing sparse or incomplete point clouds. Building on this, Yu et al. introduced \textbf{PoinTr} \cite{yu2021pointr}, a point cloud completion network based on the Transformer architecture. By introducing geometry-aware self-attention, PoinTr dynamically captures global context and local structural relationships between discrete point sets, thereby more accurately inferring and restoring the fine topology and 3D geometric features of missing regions.

These works strongly indicate that utilizing GNNs to capture the local topology of unstructured data and regress missing attributes is a highly promising technical route. Inspired by these advancements, and considering the over-parameterized nature of 3DGS, this paper introduces a lightweight two-layer GNN with hash-context priors at the decoder side to complement interpolation-based restoration of missing 3D Gaussian attributes.

\section{Method}
\label{sec:method}

\begin{figure*}[htbp]
    \centering
    \includegraphics[width=\textwidth]{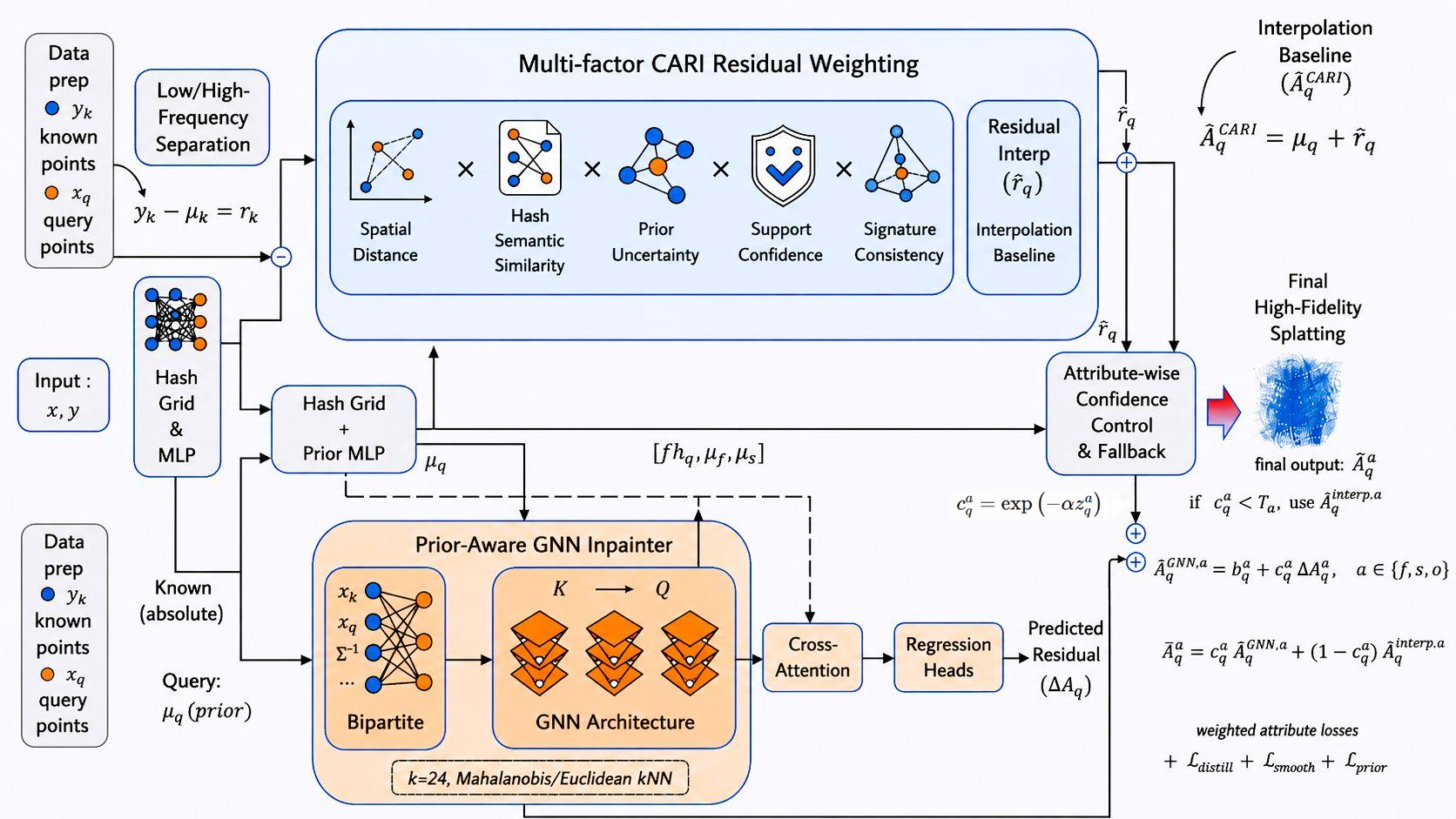} 
    \Description{Architecture diagram of the decoder-side dual branch error concealment method, with an upper CARI interpolation branch, a lower EdgeConv GNN branch, cross-attention fusion with hash-grid priors, and confidence-gated attribute reconstruction.}
    \caption{Architecture of the proposed prior-aware dual-branch error concealment pipeline at the decoder side. 
\textbf{Upper branch (CARI):} Known anchor attributes are decomposed into prior predictions $\boldsymbol{\mu}_k^a$ and branch-wise residuals $\mathbf{r}_k^a{=}\mathbf{y}_k^a{-}\boldsymbol{\mu}_k^a$ via the decoded Hash Grid and Prior MLP. Neighbor residuals are aggregated through multi-factor weights combining spatial proximity, semantic similarity, prior uncertainty, support confidence, and signature consistency, producing the interpolation baseline $\mathbf{A}_q^{CARI}$. 
\textbf{Lower branch (GNN):} A bipartite graph between known nodes (absolute attributes) and query nodes (prior predictions $\boldsymbol{\mu}_q$) is processed by a lightweight two-layer EdgeConv backbone. A Cross-Attention Fusion module integrates the hash-grid prior $[\mathbf{f}_q^h, \boldsymbol{\mu}_f, \boldsymbol{\mu}_s]$, after which attribute-specific regression heads produce confidence-gated residuals to form $\mathbf{A}_q^{GNN,a}{=}\mathbf{b}_q^a{+}c_q^a\Delta\mathbf{A}_q^a$. 
The final output is controlled by attribute-wise confidence, with fallback to interpolation when the confidence of an attribute is insufficient. The model is trained with weighted attribute reconstruction losses, interpolation distillation, graph smoothness, and prior regularization.}
    \label{fig:architecture}
\end{figure*}

In this section, we present a robust 3DGS transmission framework. Let the compressed 3DGS model be represented by a set of $N$ anchors $\mathcal{A} = \{a_i\}_{i=0}^{N-1}$, where each anchor $a_i$ encapsulates a feature vector $\mathbf{f}_i$, scaling factors $\mathbf{s}_i$, learnable offsets $\mathbf{o}_i$, and an existence mask $m_i$. To enhance resilience against network impairments, we partition the bitstream into spatial blocks and $M$ independent interleaving lanes, yielding packet units $\mathcal{P}_{t,\lambda}$.

The proposed robust transmission and error concealment framework aims to address the quality degradation of 3D Gaussian Splatting (3DGS) compressed representations under unreliable network conditions. As illustrated in Fig.~\ref{fig:framework}, the system primarily consists of two phases. The detailed decoder-side architecture is further depicted in Fig.~\ref{fig:architecture}.

\begin{enumerate}
    \item \textbf{Encoding and Transmission Phase:} We introduce \textit{Anchor-level Atomic Packaging} and \textit{Stratified Random Grouping} strategies. These strategies aim to reshape the failure mode of packet loss, transforming "artifacts caused by attribute inconsistency" into "spatially uniform loss of complete anchors."
    \item \textbf{Decoding and Repair Phase:} The decoder first localizes missing packets and defines the anchors within them as the set to be repaired. It then applies Context-Aware Residual Interpolation (CARI) as a robust interpolation baseline that uses hash-grid priors and neighboring residuals. On top of this baseline, a lightweight two-layer Graph Neural Network (GNN)---augmented with Cross-Attention Fusion over hash-grid priors---is used to refine high-confidence attribute residuals, achieving confidence-controlled end-to-end error concealment.
\end{enumerate}

Additionally, to evaluate the effectiveness of the GNN repair, we implement a \textit{k}-Nearest Neighbor (k-NN) based weighted interpolation module as a comparative baseline.

\subsection{Preliminaries: 3D Gaussian Splatting}
\label{subsec:preliminaries_3dgs}

3D Gaussian Splatting (3DGS) \cite{kerbl20233d} represents a 3D scene using a set of anisotropic Gaussians. Each Gaussian is defined by a center $\mu \in \mathbb{R}^3$ and a covariance matrix $\Sigma$, evaluated as:
\begin{equation}
    G(x) = e^{-\frac{1}{2}(x-\mu)^T\Sigma^{-1}(x-\mu)}
\end{equation}
To ensure positive semi-definiteness, $\Sigma$ is parameterized by a scaling matrix $S$ and a rotation matrix $R$ via $\Sigma = R S S^T R^T$. Additionally, each Gaussian carries an opacity $\alpha$ and view-dependent color represented by Spherical Harmonics (SH).

For rendering, 3D Gaussians are projected onto the 2D image plane. The pixel color $C$ is computed via front-to-back $\alpha$-blending of $N$ sorted Gaussians overlapping the pixel:
\begin{equation}
    C = \sum_{i \in \mathcal{N}} c_i \alpha'_i \prod_{j=1}^{i-1} (1 - \alpha'_j)
\end{equation}
where $c_i$ is the color and $\alpha'_i$ is the final opacity combining $\alpha$ with the 2D Gaussian evaluation. This explicit representation enables high-quality real-time rendering but is sensitive to parameter inconsistencies caused by packet loss.

To mitigate the substantial storage overhead of vanilla 3DGS, Scaffold-GS \cite{lu2024scaffold} introduces a structured, anchor-based representation. Unlike the explicit storage of millions of individual Gaussians, Scaffold-GS utilizes a sparse set of \textit{anchors} initialized from Structure-from-Motion (SfM) points. Each anchor is associated with a feature vector, a scaling factor, and a set of learnable offsets. During the rendering phase, a lightweight Multi-Layer Perceptron (MLP) dynamically decodes these anchor attributes to spawn multiple 3D Gaussians, predicting their opacity, color, rotation, and final scaling. This hierarchical "Anchor-to-Gaussian" design exploits local redundancies, significantly reducing the number of explicit parameters while maintaining high-fidelity rendering.

Building upon this architecture, HAC++ \cite{chen2025hac} further advances compression efficiency by exploiting the mutual information between unorganized anchors and a structured feature grid. HAC++ introduces a Hash-grid Assisted Context (HAC) framework, where a compact, binarized 3D Hash Grid is queried to assist in modeling the context of anchor attributes. Specifically, the interpolated features from the hash grid are used to predict the probability distributions of the anchor parameters (features, scales, and offsets) for arithmetic coding. By combining this context modeling with adaptive quantization and mask-based pruning, HAC++ achieves a storage reduction of over $100\times$ compared to vanilla 3DGS. In this work, we adopt the compressed bitstream structure of HAC++ as the baseline for our transmission and error concealment study.
\subsection{Encoder-Side Strategy}
\label{subsec:encoder_strategy}

\subsubsection{Anchor-level Atomic Packaging}
\label{subsec:atomic_packaging}

To maximize entropy coding efficiency, existing 3DGS compression frameworks (e.g., HAC++ \cite{chen2025hac}) predominantly adopt an \textit{Attribute-Decoupled Packaging} strategy. This approach independently clusters and encodes the features, scaling factors, and offsets of all anchors to exploit their respective statistical redundancies. However, such a strategy overlooks the \textit{intrinsic coupling} among anchor attributes. In lossy transmission or packet loss scenarios, the loss of a single attribute stream results in the reconstruction of "Corrupted Anchors" characterized by inconsistent geometric and appearance parameters.

To validate this hypothesis, we conducted a comparative experiment investigating reconstruction quality under two distinct packet loss patterns:

\begin{itemize}
    \item \textbf{Atomic Loss (Joint Dropping)}, where all attribute data corresponding to 20\% of the anchors are randomly discarded. In this scenario, the affected anchors are treated as \textit{invalid} and excluded from the rendering computation. Their spatial coordinates are retained solely as \textit{indices} to guide the subsequent joint recovery via neighborhood interpolation.
    
    \item \textbf{Partial Loss (Attribute Erosion)}, where the anchors remain \textit{active} and participate in the rendering process. However, a specific attribute (e.g., scaling or offset) is randomly dropped for 20\% of the anchors, and the missing values are filled via interpolation.
\end{itemize}

\begin{table*}[t]
\centering
\caption{Motivation experiment for anchor-level atomic packaging under 20\% random loss and interpolation-based concealment. ``Packed + Interp.'' drops 20\% complete packed anchors and then interpolates their attributes, while the other lossy columns drop only one native HAC++ attribute stream for 20\% of anchors and interpolate the missing attribute. Parenthesized values are PSNR drops relative to the normal reference in this table. For consistency with the main packet-loss evaluation, the normal and packed-anchor columns use the later dataset-average protocol, and attribute-loss drops are recomputed against these normal references. T\&T denotes Tanks \& Temples.}
\label{tab:atomic_packaging_motivation}
\resizebox{\textwidth}{!}{
\begin{tabular}{@{}lcccccc@{}}
\toprule
\textbf{Dataset} & \textbf{Normal} & \textbf{Packed + Interp.} & \textbf{Drop Feat} & \textbf{Drop Scale} & \textbf{Drop Offset} & \textbf{Worst Attr.} \\
\midrule
BungeeNeRF & 28.5042 & 23.7115 (-4.7927) & 22.0972 (-6.4070) & 23.4065 (-5.0977) & 20.7454 (-7.7588) & Offset \\
Mip-NeRF 360 & 29.1046 & 25.7325 (-3.3721) & 23.0190 (-6.0856) & 24.8911 (-4.2135) & 24.3487 (-4.7559) & Feat \\
T\&T & 24.1378 & 22.2131 (-1.9247) & 19.7191 (-4.4187) & 20.2571 (-3.8807) & 20.9438 (-3.1940) & Feat \\
\midrule
All & 28.4446 & 24.7843 (-3.6603) & 22.3103 (-6.1343) & 23.8146 (-4.6300) & 22.5600 (-5.8846) & Feat \\
\bottomrule
\end{tabular}
}
\end{table*}

Table~\ref{tab:atomic_packaging_motivation} provides a compact empirical motivation for atomic packaging. Under the same 20\% random loss rate and interpolation repair strategy, dropping complete packed anchors produces a smaller all-scene PSNR drop (3.6603~dB) than independently losing feature, scale, or offset streams (6.1343~dB, 4.6300~dB, and 5.8846~dB, respectively). Feature loss is the worst case on Mip-NeRF 360, T\&T, and the all-scene average, while offset loss is most destructive on BungeeNeRF. This indicates that retaining anchors with internally inconsistent attributes is more harmful than treating the corresponding anchors as completely missing. Jointly packaging correlated anchor attributes therefore avoids anchor-internal mismatch and improves reconstruction stability after interpolation-based error concealment.

In response to these findings, we propose the \textit{Atomic Packaging} strategy. We strictly bind all attributes of each anchor $a_i$ into an indivisible tuple and encapsulate them into the same transmission packet $\mathcal{P}_j$. Formally, the $j$-th packet is defined as:

\begin{equation}
    \mathcal{P}_j = \{ (\mathbf{f}_i, \mathbf{s}_i, \mathbf{o}_i, m_i) \mid i \in \mathcal{G}_j \}
\end{equation}

where $\mathcal{G}_j$ denotes the set of anchor indices contained within the $j$-th packet, and $m_i$ is the existence mask. Here, $\mathcal{P}_j$ is a generic packet notation; in the stratified implementation introduced next, the packet index $j$ is instantiated by a block-lane pair $(t,\lambda)$, yielding the concrete packet $\mathcal{P}_{t,\lambda}$. Mechanically, this strategy enforces an ``All-or-Nothing'' degradation mode for channel packet loss. If a specific packet $\mathcal{P}_j$ is lost, all anchors contained within it are treated as completely missing. This design effectively eliminates attribute inconsistency, thereby simplifying the intractable problem of removing complex visual artifacts into a well-defined \textit{Point Cloud Completion} task.

\subsubsection{Stratified Random Grouping}
\label{subsec:uniform_grouping}

To combat burst packet loss in network transmission, we fundamentally alter the conventional adjacency-based grouping strategy derived from Morton code (Z-order). While Morton ordering preserves spatial locality, it becomes a liability under burst loss conditions: spatially contiguous blocks of anchors are lost simultaneously, creating large-scale spatial voids that are intractable to repair.

To address this while preserving the capability for streaming encoding, we instantiate \textit{Stratified Random Grouping} as a local multi-lane interleaving procedure. Unlike standard approaches that simply segment the Morton-ordered sequence into sequential packets, we introduce a randomized interleaving mechanism. Let $r_i \in \{0,\dots,N-1\}$ denote the zero-based rank of anchor $a_i$ in the Morton-sorted sequence, let $N_{\mathrm{blk}}$ denote the maximum number of anchors per spatial block, and let $\xi$ be the shared packetization seed. The number of spatial blocks is $S=\lceil N/N_{\mathrm{blk}}\rceil$. The construction process is as follows:

\begin{enumerate}
    \item \textbf{Spatial Blocking.} First, the Morton-sorted anchor sequence is divided into $S$ sequential spatial blocks, denoted as $\{\mathcal{B}_0, \mathcal{B}_1, \dots, \mathcal{B}_{S-1}\}$. Each block contains at most $N_{\mathrm{blk}}$ spatially adjacent anchors, corresponding to the \texttt{max\_batch\_size} used by the implementation:
    \begin{equation}
        t_i = \left\lfloor \frac{r_i}{N_{\mathrm{blk}}} \right\rfloor .
    \end{equation}
    where $t_i \in \{0,\dots,S-1\}$ is the spatial block (or streaming step) index of anchor $a_i$.
    
    \item \textbf{Hash Interleaving.} Within each spatial block $\mathcal{B}_{t_i}$, we employ a pseudo-random hash function (SplitMix64) to uniformly distribute anchors into $M$ parallel transmission \textit{lanes}:
    \begin{equation}
        \ell_i = \operatorname{SplitMix64}(r_i \oplus \xi) \bmod M .
    \end{equation}
    where $\ell_i \in \{0,\dots,M-1\}$ is the lane index of anchor $a_i$, and $\oplus$ denotes bitwise XOR. The hash is evaluated on the global Morton rank $r_i$, matching the encoder-side interleaver.
    
    \item \textbf{Packet Construction.} The final transmission packet $\mathcal{P}_{t,\lambda}$ is constructed by aggregating anchors from block $\mathcal{B}_t$ that are assigned to lane $\lambda$:
    \begin{equation}
        \mathcal{P}_{t,\lambda} =
        \{(\mathbf{f}_i,\mathbf{s}_i,\mathbf{o}_i,m_i)
        \mid t_i=t,\; \ell_i=\lambda\}.
    \end{equation}
    where $\lambda$ denotes the lane identifier and is distinct from the existence mask $m_i$. Equivalently, the scalar packet identifier used during training is $j=tM+\lambda$.
\end{enumerate}

The anchor coordinates, hash-grid parameters, and prior MLP weights are transmitted as global side information as in HAC-style decoding, while each packet carries the recoverable per-anchor attributes. In this work, the global side information is treated as protected control metadata and is assumed to be delivered reliably through retransmission, forward error correction, or a reliable control channel; the packet-loss simulation focuses on erasures of the per-anchor attribute \texttt{.pak} payloads. Under this strategy, each packet $\mathcal{P}_{t,\lambda}$ carries approximately $1/M$ of the information content for the specific spatial region covered by $\mathcal{B}_t$. In the event of packet loss, even if a specific packet $\mathcal{P}_{t,\lambda}$ is dropped, the corresponding spatial region retains $(M-1)/M$ of its anchors transmitted via other lanes. Consequently, burst loss manifests as a uniform reduction in local point density (i.e., \textit{spatial downsampling}) rather than a complete geometric absence. This preservation of local geometric structure provides reliable support points for the subsequent neighborhood-based restoration.

\subsection{Context-Aware Residual Interpolation (CARI)}
\label{subsec:baseline}

To fully leverage the prior information embedded in the HAC++ coding structure, we design an enhanced interpolation algorithm that serves as both a standalone baseline and the robust low-frequency branch in our dual-branch pipeline (upper branch of Fig.~\ref{fig:architecture}). Unlike traditional interpolation methods that rely solely on geometric distances, this approach exploits the semantic similarity of Hash Grid features and the \textit{prior} prediction capability of MLPs to guide the recovery process. The method consists of three key steps: Prior Prediction, Joint Weighting, and Residual Compensation.

\paragraph{Prior Prediction Estimation}
For every anchor (whether known $k \in \mathcal{K}$ or missing $q \in \mathcal{Q}$), we utilize the decoded Hash Grid features $\mathbf{f}^h$ and a shared MLP to obtain a prior attribute prediction. This value is the predicted center of the HAC++ conditional distribution, and is used in our restoration pipeline as a baseline estimate rather than as the final recovered attribute. Let $\mathbf{x}$ be the anchor coordinate; we obtain this prior prediction $\boldsymbol{\mu}$ via the Hash encoder and a grid MLP:

\begin{equation}
    \boldsymbol{\mu}(\mathbf{x}) = \text{MLP}_{\text{grid}}(\text{HashEncoder}(\mathbf{x}))
\end{equation}

where $\boldsymbol{\mu}(\mathbf{x})$ encompasses the prior predictions for features, scales, and offsets. For each attribute branch $a \in \{f,s,o\}$ of a known anchor $k$, let $\mathbf{y}_k^a$ denote the actual decoded attribute and $\boldsymbol{\mu}_k^a$ denote the corresponding prior prediction. The branch-wise residual is defined as:
\begin{equation}
    \mathbf{r}_k^a = \mathbf{y}_k^a - \boldsymbol{\mu}_k^a .
\end{equation}
When the same operation applies to all branches, we omit the superscript $a$ for compact notation. This step strips away the low-frequency commonalities of the scene, allowing the subsequent interpolation to focus on recovering high-frequency detailed residuals.

\paragraph{Geometric-Semantic Joint Weighting}
For each missing anchor $q$, we first collect a Euclidean candidate set $\mathcal{C}(q)$ with size $L$ from the known set $\mathcal{K}$, and then retain the strongest $K$ neighbors after multi-factor scoring. In the implementation we use $L{=}128$ and $K{=}8$ by default. For a candidate neighbor $k$, the basic similarity and reliability terms are
\begin{align}
    d_{qk} &= \|\mathbf{x}_q-\mathbf{x}_k\|_2, \\
    h_{qk} &= \left(\frac{\operatorname{CosSim}(\mathbf{f}_q^h,\mathbf{f}_k^h)+1}{2}\right), \\
    \rho_{qk} &= \left(\frac{\operatorname{CosSim}(\mathbf{g}_q,\mathbf{g}_k)+1}{2}\right).
\end{align}
where $d_{qk}$ is the spatial distance, $h_{qk}$ is the hash-context similarity, and $\rho_{qk}$ measures local residual-pattern consistency using the compact signature $\mathbf{g}$. Following the implementation, we form five scalar scores: $w^{geo}_{qk}{=}(d_{qk}+\epsilon)^{-\alpha/2}$, $w^{ctx}_{qk}{=}h_{qk}^{\beta}$, $w^{unc}_{qk}{=}(\sigma_k+\epsilon)^{-\gamma}$, $w^{conf}_{qk}{=}c_k$, and $w^{sig}_{qk}{=}\rho_{qk}^{\delta}$. The unnormalized interpolation score is then simply
\begin{equation}
    u_{qk} =
    w^{geo}_{qk}
    w^{ctx}_{qk}
    w^{unc}_{qk}
    w^{conf}_{qk}
    w^{sig}_{qk}.
\end{equation}
where $\sigma_k$ is the attribute-wise prior scale or uncertainty of the neighbor, and $c_k$ indicates the reliability of the neighbor support. In other words, a neighbor receives a higher score when it is spatially close, has a similar hash-grid context, has lower prior uncertainty, comes from a reliable decoded support, and exhibits a similar local residual signature. The normalized weight is
\begin{equation}
    \bar{w}_{qk} =
    \frac{u_{qk}}
    {\sum_{j\in \mathcal{N}(q)} u_{qj}+\epsilon}.
\end{equation}
This formulation matches the implemented CARI branch: spatial proximity, hash-feature similarity, prior uncertainty, available support confidence, and signature consistency jointly determine which residuals are trusted.

\paragraph{Implementation Constants}
Unless otherwise specified, the decoder uses a candidate pool of $L{=}128$ nearest anchors and keeps the strongest $K{=}8$ supports after joint weighting. The default CARI exponents are $\alpha{=}1.0$, $\beta{=}2.0$, $\gamma{=}1.0$, and $\delta{=}1.5$. The support confidence is set to $c_k{=}1$ for attributes already decoded successfully, and to $c_k{=}0.5$ for attributes recovered by concealment in a previous refinement pass. The local signature $\mathbf{g}$ is computed from compact residual statistics: five feature-residual means, one scale-residual mean, and one offset-residual mean. The prior uncertainty $\sigma_k$ is obtained from the HAC++ grid MLP scales; in practice we use the corresponding attribute-wise scalar uncertainty, i.e., the feature-channel scale for feature groups, the mean scale uncertainty for scaling, and the mean offset uncertainty for offsets. We run two CARI refinement iterations by default; the optional ridge refinement is disabled unless explicitly enabled.

\paragraph{Residual Interpolation and Reconstruction}
We perform weighted interpolation on the residuals $\mathbf{r}_k$ within the local neighborhood $\mathcal{N}(q)$ to estimate the residual $\hat{\mathbf{r}}_q$ for the missing point:

\begin{equation}
    \hat{\mathbf{r}}_q = \sum_{k \in \mathcal{N}(q)} \bar{w}_{qk} \mathbf{r}_k .
\end{equation}

The final CARI estimate combines the query prior prediction with the interpolated residual:
\begin{equation}
    \hat{\mathbf{A}}_q^{CARI} = \boldsymbol{\mu}_q + \hat{\mathbf{r}}_q .
\end{equation}
If a query has no valid known support for an attribute, the branch falls back to the prior prediction for that attribute. For the feature branch, the implementation can additionally apply a lightweight ridge-refinement step to the interpolated residuals.

\subsection{Prior-Aware GNN Residual Inpainting}
\label{subsec:gnn_inpainting}

At the decoder side, based on the reception status of packets, the global set of anchors is partitioned into two subsets: the \textit{Known Set} $\mathcal{K}$ (successfully decoded anchors) and the \textit{Query Set} $\mathcal{Q}$ (missing anchors due to packet loss). As illustrated in the lower branch of Fig.~\ref{fig:architecture}, we employ a lightweight two-layer Graph Neural Network (GNN) to robustly recover the attributes of $\mathcal{Q}$ from $\mathcal{K}$. The pipeline proceeds in five stages: bipartite graph construction, asymmetric node feature encoding, message passing with cross-attention fusion, residual regression, and attribute-wise confidence control.

\paragraph{Anisotropic Bipartite Graph Construction}
For each query anchor $q \in \mathcal{Q}$, we construct a local bipartite graph $\mathcal{G} = (\mathcal{Q}, \mathcal{K}, \mathcal{E})$ by searching its $k$-nearest neighbors ($k{=}24$) within the known set $\mathcal{K}$. When anisotropic covariance information is available for the known anchors, we use a Mahalanobis distance defined by each candidate support anchor:
\begin{equation}
    D_{\text{mahal}}(\mathbf{x}_q, \mathbf{x}_k) =
    \sqrt{(\mathbf{x}_q-\mathbf{x}_k)^T
    \boldsymbol{\Sigma}_k^{-1}
    (\mathbf{x}_q-\mathbf{x}_k)} .
\end{equation}
where $\boldsymbol{\Sigma}_k = \mathbf{R}_k \mathbf{S}_k \mathbf{S}_k^T \mathbf{R}_k^T$ is constructed from the known anchor's scale $\mathbf{S}_k$ and rotation $\mathbf{R}_k$. If this covariance is unavailable or incompatible with the current subset, the implementation falls back to Euclidean $k$-NN. The edge set $\mathcal{E}$ contains directed edges pointing from $\mathcal{K}$ to $\mathcal{Q}$:
\begin{equation}
    \mathcal{N}(q) =
    \underset{k \in \mathcal{K}}{\text{top-}k} \;\;
    D(\mathbf{x}_q,\mathbf{x}_k),
\end{equation}
where $D$ denotes the Mahalanobis or Euclidean distance used by the current configuration, and $\mathbf{x}$ denotes the world coordinates of the anchor.

\paragraph{Node Feature Input and Interpolation Prior}
The input features are designed asymmetrically based on the node type:
\begin{itemize}
    \item \textbf{Source Nodes ($k \in \mathcal{K}$):} The input concatenates the fully decoded attributes: position $\mathbf{x}_k$, hash-grid feature $\mathbf{f}_k^h$, anchor feature $\mathbf{f}_k$, scale $\mathbf{s}_k$, offset $\mathbf{o}_k$, and existence mask $m_k$.
    \item \textbf{Target Nodes ($q \in \mathcal{Q}$):} Since the true attribute values are unavailable, the input substitutes them with MLP prior predictions: position $\mathbf{x}_q$, hash-grid feature $\mathbf{f}_q^h$, prior predictions $(\boldsymbol{\mu}_f, \boldsymbol{\mu}_s, \boldsymbol{\mu}_o)$ obtained via $\text{MLP}_{\text{grid}}(\text{HashEnc}(\mathbf{x}_q))$, and a zero existence mask. Additionally, an IDW interpolation result $\mathbf{A}_q^{interp}$, using Mahalanobis or Euclidean neighbors according to covariance availability, is concatenated to the query node features, providing a strong low-frequency structural initialization.
\end{itemize}

\paragraph{Multi-layer GNN, Cross-Attention, and Residual Prediction}
The GNN backbone consists of two residual message-passing layers, each employing an EdgeConv operator \cite{wang2019dynamic} with PreNorm and skip connections. At each layer $l\in\{0,1\}$, node embeddings are updated via:
\begin{equation}
    \mathbf{h}_i^{(l+1)} = \mathbf{h}_i^{(l)} + \text{EdgeConv}^{(l)}\!\bigl(\text{LN}(\mathbf{h}_i^{(l)}),\, \{\text{LN}(\mathbf{h}_j^{(l)})\}_{j \in \mathcal{N}(i)}\bigr)
\end{equation}
where $\text{LN}$ denotes Layer Normalization.

After the final message-passing layer, the query node representations $\mathbf{h}_q$ are refined through a \textit{Cross-Attention Fusion} module (see the ``Cross-Attention'' block in Fig.~\ref{fig:architecture}) that adaptively integrates the hash-grid prior:
\begin{equation}
    \mathbf{h}_q' = \text{CrossAttn}\!\bigl(Q{=}\mathbf{h}_q,\;\; KV{=}[\mathbf{f}_q^h,\, \boldsymbol{\mu}_f,\, \boldsymbol{\mu}_s]\bigr)
\end{equation}
where $\mathbf{f}_q^h$ is the hash-grid feature and $\boldsymbol{\mu}_f, \boldsymbol{\mu}_s$ are the MLP prior predictions for feature and scale, respectively. This mechanism enables the network to dynamically balance between neighborhood aggregation cues and hash-grid priors on a per-query basis.

The fused representations are then passed through attribute-specific regression heads (MLPs) to produce residuals for feature, scale, and offset, together with a mask logit:
\begin{equation}
    \{ \Delta \mathbf{f}_q, \Delta \mathbf{s}_q, \Delta \mathbf{o}_q, \hat{m}_q \} = \text{Heads}(\mathbf{h}_q')
\end{equation}
For each continuous attribute $a\in\{f,s,o\}$, the final GNN prediction is obtained by adding a confidence-gated residual to an attribute baseline $\mathbf{b}_q^a$:
\begin{equation}
    \hat{\mathbf{A}}_q^{GNN,a}
    =
    \mathbf{b}_q^a + c_q^a \Delta \mathbf{A}_q^a .
\end{equation}
In the default residual-on-prior configuration, $\mathbf{b}_q^a=\boldsymbol{\mu}_q^a$; when residual interpolation is enabled, $\mathbf{b}_q^a$ can instead be the corresponding IDW/CARI baseline. The scalar confidence $c_q^a \in [0,1]$ is derived from the average z-score of the residual for attribute $a$:
\begin{equation}
    z_q^a =
    \frac{1}{d_a}
    \sum_{j=1}^{d_a}
    \left|
    \frac{\Delta A_{q,j}^{a}}{\sigma_{q,j}^{a}+\epsilon}
    \right|,
    \qquad
    c_q^a = \exp(-\alpha z_q^a).
\end{equation}
where $d_a$ is the dimensionality of attribute $a$, $\sigma_{q,j}^{a}$ is the $j$-th prior scale, and $\alpha$ controls the gating sharpness. This residual-on-baseline formulation reduces the mapping difficulty by constraining the network to predict only local deviations while automatically suppressing unreliable attributes. The mask logit $\hat{m}_q$, after Sigmoid activation, determines whether the recovered anchor participates in the final rendering.

\paragraph{Attribute-wise Confidence Control and Fallback}
While the GNN module efficiently recovers high-frequency details, it may occasionally produce unstable estimations in severely ill-posed regions lacking local context. To guarantee rendering stability, we apply attribute-wise confidence control to the GNN output.

Let $\hat{\mathbf{A}}_q^{interp,a}$ denote the interpolation baseline for attribute $a$, obtained from CARI or the IDW branch used by the current configuration. The confidence $c_q^a$ defined above is reused as the attribute-wise blending coefficient:
\begin{equation}
    \bar{\mathbf{A}}_q^a =
    c_q^a \hat{\mathbf{A}}_q^{GNN,a}
    +(1-c_q^a)\hat{\mathbf{A}}_q^{interp,a}.
\end{equation}
This soft blend provides a candidate output, but it is not used blindly. The confidence value is a reliability indicator derived from the prior-normalized residual magnitude, rather than a perfectly calibrated linear quality score. Therefore, even a small nonzero $c_q^a$ may inject an unstable GNN residual into geometry-sensitive attributes. We then apply a hard-thresholded fallback per attribute as a safety switch:
\begin{equation}
    \tilde{\mathbf{A}}_q^a =
    \begin{cases}
    \hat{\mathbf{A}}_q^{interp,a}, & c_q^a < T_a, \\
    \bar{\mathbf{A}}_q^a, & c_q^a \ge T_a .
    \end{cases}
\end{equation}
Thus, Eq.~(23) defines how to blend GNN and interpolation when the GNN branch is considered reliable, while Eq.~(24) decides whether that blended candidate should be trusted at all. This mechanism does not discard all GNN predictions with a single global score. Instead, each attribute is judged independently. For geometry-sensitive attributes such as scale and offset, the implementation can fall back to interpolation when their confidence is not sufficiently high, while high-confidence predictions are still blended into the final attributes. Feature attributes can therefore participate more actively when their residual confidence is reliable.

In the final decoder configuration, residuals are clipped to $\tau{=}3.0$ prior standard deviations and the confidence sharpness is set to $\alpha{=}0.5$. The fallback thresholds are attribute-specific: $T_f{=}0.10$ for features, $T_s{=}1.01$ for scales, and $T_o{=}1.01$ for offsets and masks. Since $c_q^a\leq 1$, this conservative feature-focused setting keeps geometry-sensitive scale and offset predictions on the interpolation branch by default, while allowing only high-confidence feature residuals to be softly blended. More generally, all feature, scale, and offset attributes are subject to the same confidence-controlled fallback rule; the reported configuration simply chooses a stricter threshold for geometry-sensitive attributes. This choice reflects the empirical observation that CARI can be stronger in smooth or small scenes, whereas the GNN branch is most useful for high-frequency appearance details.

\paragraph{Self-supervised Decoder-side Training Strategy}
To optimize the parameters $\phi$ of the GNN module without relying on additional annotated data or encoder-side co-training, we adopt an offline scene-specific ``Pseudo-dropout'' calibration mechanism. In our experimental pipeline, the complete locally stored compressed scene is first decoded before packet-loss simulation, using the available bitstreams and HAC++ MLP checkpoint. This decoded scene is used only as self-supervision for local GNN calibration: packet groups are pseudo-dropped at a 20\% rate, the dropped anchors become queries, and the complete local attributes serve as reconstruction targets. During the subsequent packet-loss evaluation, we simulate missing \texttt{.pak} packets, load the trained scene-specific weights, and perform no online optimization or access to the attributes of the dropped packets. The main optimization is performed in the parameter domain, and an optional rendering loss can be enabled for validation-oriented fine tuning. The hybrid loss function (annotated on the right side of Fig.~\ref{fig:architecture}) is formulated as:
\begin{equation}
    \mathcal{L}=\mathcal{L}_{rec}+\mathcal{L}_{reg}.
\end{equation}
where the reconstruction and regularization terms are
\[
\begin{aligned}
    \mathcal{L}_{rec} ={}&
    w_f\mathcal{L}_{f}
    +w_s\mathcal{L}_{s}
    +w_o\mathcal{L}_{o}
    +w_m\mathcal{L}_{m},\\
    \mathcal{L}_{reg} ={}&
    \lambda_{distill}\mathcal{L}_{distill}
    +\lambda_{smooth}\mathcal{L}_{smooth}\\
    &\quad
    +\lambda_{prior}\mathcal{L}_{prior}
    +\lambda_{render}\mathcal{L}_{render}.
\end{aligned}
\]
In the following definitions, $q$ ranges over pseudo-dropped query anchors in the current training mini-batch $\mathcal{B}_Q \subset \mathcal{Q}$:
\begin{itemize}
    \item $\mathcal{L}_{f} = \sum_q \|\hat{\mathbf{f}}_q-\mathbf{f}_q^{gt}\|_1$ and $\mathcal{L}_{s} = \sum_q \|\hat{\mathbf{s}}_q-\mathbf{s}_q^{gt}\|_1$ supervise feature and scale reconstruction.
    \item $\mathcal{L}_{o} = \sum_q \operatorname{Charb}(\hat{\mathbf{o}}_q-\mathbf{o}_q^{gt})$ supervises offset reconstruction with a Charbonnier penalty; the implementation can increase the offset weight on active mask entries.
    \item $\mathcal{L}_{m}$ is the mask reconstruction loss: $\mathcal{L}_{m} = \sum_q \operatorname{BCE}(\hat{m}_q,m_q^{gt})$. It supervises the recovered existence or offset-validity mask. Here $m_q^{gt}$ is the decoded mask of the pseudo-dropped anchor in the locally reconstructed full scene; it is not the packet-loss indicator.
    \item $\mathcal{L}_{distill} = \sum_{q,a} \|\hat{\mathbf{A}}_q^{GNN,a}-\hat{\mathbf{A}}_q^{interp,a}\|_1$ regularizes GNN predictions toward the interpolation teacher, which is the IDW/CARI baseline used by the current configuration.
    \item $\mathcal{L}_{smooth}$ is a local query-query graph regularizer. Let $\mathcal{E}_Q=\{(q,q')\mid q'\in\operatorname{kNN}_{k_{\mathrm{lap}}}(\mathbf{x}_q;\mathcal{B}_Q\setminus\{q\})\}$ denote the KNN graph built only among query anchors in the current mini-batch. The implementation uses
    \[
    \mathcal{L}_{smooth} =
    \frac{1}{|\mathcal{E}_Q|}
    \sum_{(q,q')\in\mathcal{E}_Q}
    \omega_{qq'}
    \sum_{a\in\mathcal{S}}
    \|\hat{\mathbf{A}}_q^a-\operatorname{sg}(\hat{\mathbf{A}}_{q'}^a)\|_1 ,
    \]
    where $\mathcal{S}$ is the set of selected trained attributes, $\operatorname{sg}(\cdot)$ denotes stop-gradient, and $\omega_{qq'}$ is a distance weight, set to one by default or to a Gaussian weight when Laplacian smoothing bandwidth is enabled. Thus, both $q$ and $q'$ are pseudo-dropped query anchors, while $q'$ is the local neighbor of $q$ in the training mini-batch.
    \item $\mathcal{L}_{prior} = \sum_{q,a}\frac{1}{d_a}\left\|\frac{\hat{\mathbf{A}}_q^a-\boldsymbol{\mu}_q^a}{\boldsymbol{\sigma}_q^a+\epsilon}\right\|_1$ penalizes large z-score deviations from the hash-grid prior.
    \item $\mathcal{L}_{render}$ is optional and is disabled when $\lambda_{render}=0$.
\end{itemize}
In our implementation, we empirically set $\lambda_{distill}=0.25$, $\lambda_{smooth}=0.05$, and $\lambda_{prior}=0.1$ in the main configuration. The final reported GNN uses a feature-focused training configuration, setting the feature reconstruction and distillation weights to 1.0 while setting the scale, offset, and mask training weights to 0. This matches the confidence-controlled inference design: scale and offset are recovered by the stronger interpolation branch unless future geometry-specific constraints make their GNN residuals reliable.

\section{Experiments}
\label{sec:Experiments}

\subsection{Experimental Setup}

\subsubsection{Datasets and Evaluation Metrics}
To comprehensively evaluate the effectiveness of the proposed robust transmission and error concealment framework, we conducted systematic experiments on three mainstream datasets for 3D scene reconstruction and Novel View Synthesis (NVS).

\begin{itemize}
    \item \textbf{BungeeNeRF Dataset}: We evaluate 8 challenging multi-scale urban scenes with large depth variation and rich high-frequency details \cite{xiangli2022bungeenerf}. These scenes stress the robustness of packet-loss concealment because missing anchors can affect both distant structures and close-range facade details.
    
    \item \textbf{Mip-NeRF 360 Dataset}: We use 7 challenging large-scale real-world scenes characterized by deep depth-of-field and complex unbounded backgrounds \cite{barron2022mip}. Following the domain standard, outdoor scenes are downsampled by a factor of 4 and indoor scenes by a factor of 2. Every 8th frame is held out as the test set to evaluate the generalization and restoration performance in complex physical environments.
    
    \item \textbf{Tanks \& Temples Dataset}: We selected representative real-world scenes (e.g., \textit{Truck}, \textit{Train}) to further test the robustness of the framework against complex geometric surface reconstruction and large-span viewpoint transitions \cite{knapitsch2017tanks}. The test set partitioning remains consistent with the Mip-NeRF 360 protocol.
\end{itemize}

\paragraph{Evaluation Metrics}
For rendering and restoration quality assessment, we employ three standard objective image quality metrics: Peak Signal-to-Noise Ratio (\textbf{PSNR}), Structural Similarity (\textbf{SSIM})~\cite{wang2004ssim}, and Learned Perceptual Image Patch Similarity (\textbf{LPIPS})~\cite{zhang2018lpips}. Specifically, PSNR reflects pixel-level absolute signal errors; SSIM measures the local spatial consistency of image structures; and LPIPS leverages deep network features to better align with the human visual system's perception of high-frequency detail loss and anomalous artifacts.

Note that the base compression scheme, HAC++ \cite{chen2025hac}, is configured with its default high-fidelity settings (e.g., Spherical Harmonic orders and feature quantization steps at their optimal defaults), ensuring that all packet loss and restoration experiments are established upon a high-quality reconstruction baseline. We also note that the rendering speed (FPS) remains largely unaffected by packet loss, as the loss of anchors reduces the rendering load; the observed FPS variation is within single-digit fluctuations.

\subsubsection{Packet Loss Simulation Protocol}
To faithfully reflect the performance of 3DGS bitstreams in unreliable network environments (e.g., connectionless UDP transmission or congested channels), we established a channel simulation model based on packet erasure. Unlike bit-flip errors at the physical layer, this work focuses on macro-level packet loss at the network layer. In this version, we report the representative \textbf{20\% Random Loss} setting, where packet units are independently removed with probability $p{=}0.20$. The protected global side information described in Section~\ref{subsec:uniform_grouping} is kept available for all methods and is not included in the random erasure process. For the proposed CARI and GNN variants, packet loss is implemented by copying the bitstream directory and deleting the selected \texttt{.pak} lane/block payloads; a missing \texttt{.pak} therefore removes complete anchors rather than individual attributes. For the original HAC++ baseline, the erasure is applied before our atomic regrouping to the native attribute-separated payloads, so the decoder may receive only a subset of the correlated feature, scale, and offset streams for affected anchors, and no decoder-side concealment is enabled. This setting is severe enough to expose the broken-anchor failure mode of attribute-separated transmission while keeping the experimental protocol compact and directly comparable across datasets.

To systematically verify the effectiveness of the proposed framework, we established the following comparison settings:

\subsubsection{Comparison Settings}

\begin{itemize}
    \item \textbf{HAC++ (Original)}: Serving as the vanilla reference baseline, this scheme retains the native attribute-independent packaging mechanism and Morton-order (Z-order) spatial arrangement of HAC++ \cite{chen2025hac}. When packet loss occurs, the decoder performs no concealment and directly renders using the residual data, which typically contains a large number of ``fragmented anchors'' with mismatched attribute parameters.
    
    \item \textbf{Ours (CARI):} The interpolation-based variant of our proposed framework. This scheme adopts our proposed anchor-level atomic packaging and stratified random grouping mechanisms, and utilizes the \textit{Context-Aware Residual Interpolation (CARI)} method described in Section~\ref{subsec:baseline} for decoder-side error concealment based on hash priors and local geometric features.
    
    \item \textbf{Ours (GNN)}: The GNN-enhanced variant of the proposed concealment framework. It integrates anchor-level atomic packaging, stratified random grouping, and a scene-specific lightweight GNN trained locally through offline pseudo-dropout calibration before packet-loss simulation. During evaluation, the decoder loads the trained GNN weights and applies attribute-wise confidence control to blend with or fall back to CARI.
\end{itemize}

\subsubsection{Implementation Details}
All experiments, including scene encoding/compression, packet loss simulation, and decoder-side 3D rendering and error concealment, were conducted on a high-performance workstation equipped with a single NVIDIA RTX 4080 GPU to ensure benchmark consistency for time-overhead evaluations (e.g., rendering FPS).

\paragraph{Packetization Parameters}
In the reported implementation, the Morton-sorted anchor sequence is processed in spatial blocks of at most $N_{\mathrm{blk}}{=}3000$ anchors, matching the \texttt{max\_batch\_size} stored in the packetization metadata. Each block is split into $M{=}8$ pseudo-random lanes using SplitMix64 with seed $\xi{=}1337$, and each non-empty step-lane pair is packed as one \texttt{pak\_lane\{lane\}\_s\{step\}.pak} transport unit. The number of spatial blocks is scene-dependent, $S=\lceil N/N_{\mathrm{blk}}\rceil$; for example, the Hollywood scene contains $N{=}369{,}941$ anchors and produces $S{=}124$ Morton blocks. The metadata file records the schema identifier, $M$, seed, block count, block size, ordering, and anchor count, enabling the decoder to reconstruct the packet-to-anchor mapping consistently.

\paragraph{Bitstream-size Overhead}
We compare the transmitted non-weight bitstream files to quantify the cost of robust packetization. Across the 19 evaluated scenes, the original HAC++ non-weight bitstreams occupy 159.70 MiB, while the proposed \texttt{.pak}-based packetization occupies 164.46 MiB, corresponding to a 4.77 MiB or 2.99\% increase. This comparison excludes model weight files such as \texttt{.pth}/\texttt{.pt}/\texttt{.ckpt}. If only the replaced attribute streams are counted, the original feature/scale/offset \texttt{.b} files occupy 139.70 MiB and the corresponding \texttt{.pak} files occupy 149.66 MiB, a 7.13\% increase. We therefore report the all-non-weight bitstream growth as the primary storage overhead, and use the attribute-only number as a diagnostic of the packaging cost itself.

\paragraph{Architecture of the GNN Repair Module}
Our GNN repair module is designed to balance expressiveness with inference efficiency. The backbone employs two residual \textbf{EdgeConv} \cite{wang2019dynamic} message-passing layers with Layer Normalization, using a hidden dimension of $d_h{=}128$. A \textbf{Cross-Attention Fusion} module with 4 heads subsequently integrates the hash-grid prior features as Key/Value. For the bipartite $k$-NN graph construction, we set $k{=}24$ and use Mahalanobis distance when compatible covariance matrices are available, otherwise falling back to Euclidean neighbors. Four independent regression heads (MLPs) produce per-attribute residuals for features, scales, offsets, and existence masks; the training weights can select feature-focused or full-attribute supervision.

\paragraph{Training Strategy and Optimization}
To eliminate dependency on external annotations, we adopt a self-supervised \textbf{Pseudo-dropout} training pipeline as an offline local calibration stage. For each scene, we first locally reconstruct the complete compressed scene before packet-loss simulation, build packet-aligned groups from the packetization metadata, and randomly mask $20\%$ of the packets during training. The masked anchors constitute the query set $\mathcal{Q}$, while the remaining anchors form the known set $\mathcal{K}$. The network is trained using the weighted hybrid loss described in Section~\ref{subsec:gnn_inpainting}, with the locally decoded full-scene attributes serving as pseudo ground truth. Training is performed for 15{,}000--20{,}000 iterations per scene using the Adam optimizer with an initial learning rate of $10^{-3}$ and cosine annealing decay. During packet-loss evaluation, packet loss is simulated by removing \texttt{.pak} files from a copied bitstream directory; the trained scene-specific weights are loaded directly, and no online optimization or lost-attribute supervision is used.

\subsection{Experimental Results and Analysis}
\label{subsec:results}
In this section, we report the quantitative reconstruction quality under the representative 20\% random packet-loss setting across three datasets. Table~\ref{tab:quantitative_results_wide} summarizes the average performance of the compared methods on three core metrics: PSNR, SSIM, and LPIPS. Averaged over the three dataset-level PSNR values, the no-packet-loss HAC++ reference is 27.25 dB and the proposed GNN configuration is 24.25 dB, corresponding to an overall average drop of 2.99 dB; the drop is larger on BungeeNeRF due to its more challenging large-scale high-frequency content.

\begin{table*}[htbp]
\centering
\caption{Quantitative comparison under 20\% random packet loss across three datasets. The best lossy-transmission results are highlighted in \textbf{bold}. All values are averaged over all scenes within each dataset.}
\label{tab:quantitative_results_wide}
\resizebox{\textwidth}{!}{
\begin{tabular}{@{}ll|ccc|ccc|ccc@{}}
\toprule
 & & \multicolumn{3}{c|}{\textbf{BungeeNeRF}} & \multicolumn{3}{c|}{\textbf{Mip-NeRF 360}} & \multicolumn{3}{c}{\textbf{Tanks \& Temples}} \\ 
\textbf{Loss Rate} & \textbf{Method} & \textbf{PSNR $\uparrow$} & \textbf{SSIM $\uparrow$} & \textbf{LPIPS $\downarrow$} & \textbf{PSNR $\uparrow$} & \textbf{SSIM $\uparrow$} & \textbf{LPIPS $\downarrow$} & \textbf{PSNR $\uparrow$} & \textbf{SSIM $\uparrow$} & \textbf{LPIPS $\downarrow$} \\ \midrule
 
0\% (No Packet Loss) & HAC++ (Reference) & 28.50 & 0.907 & 0.149 & 29.10 & 0.859 & 0.216 & 24.14 & 0.846 & 0.191 \\ \midrule
20\% & HAC++ (Original) & 19.55 & 0.610 & 0.380 & 20.05 & 0.590 & 0.405 & 15.20 & 0.520 & 0.455 \\
20\% & Ours (CARI) & 23.71 & 0.818 & 0.220 & 25.73 & 0.807 & 0.264 & 22.21 & \textbf{0.800} & \textbf{0.233} \\
20\% & Ours (GNN) & \textbf{24.31} & \textbf{0.827} & \textbf{0.205} & \textbf{26.13} & \textbf{0.811} & \textbf{0.258} & \textbf{22.32} & 0.796 & 0.237 \\ \bottomrule
\end{tabular}
}
\end{table*}

\begin{figure*}[htbp]
    \centering
    \includegraphics[width=\textwidth]{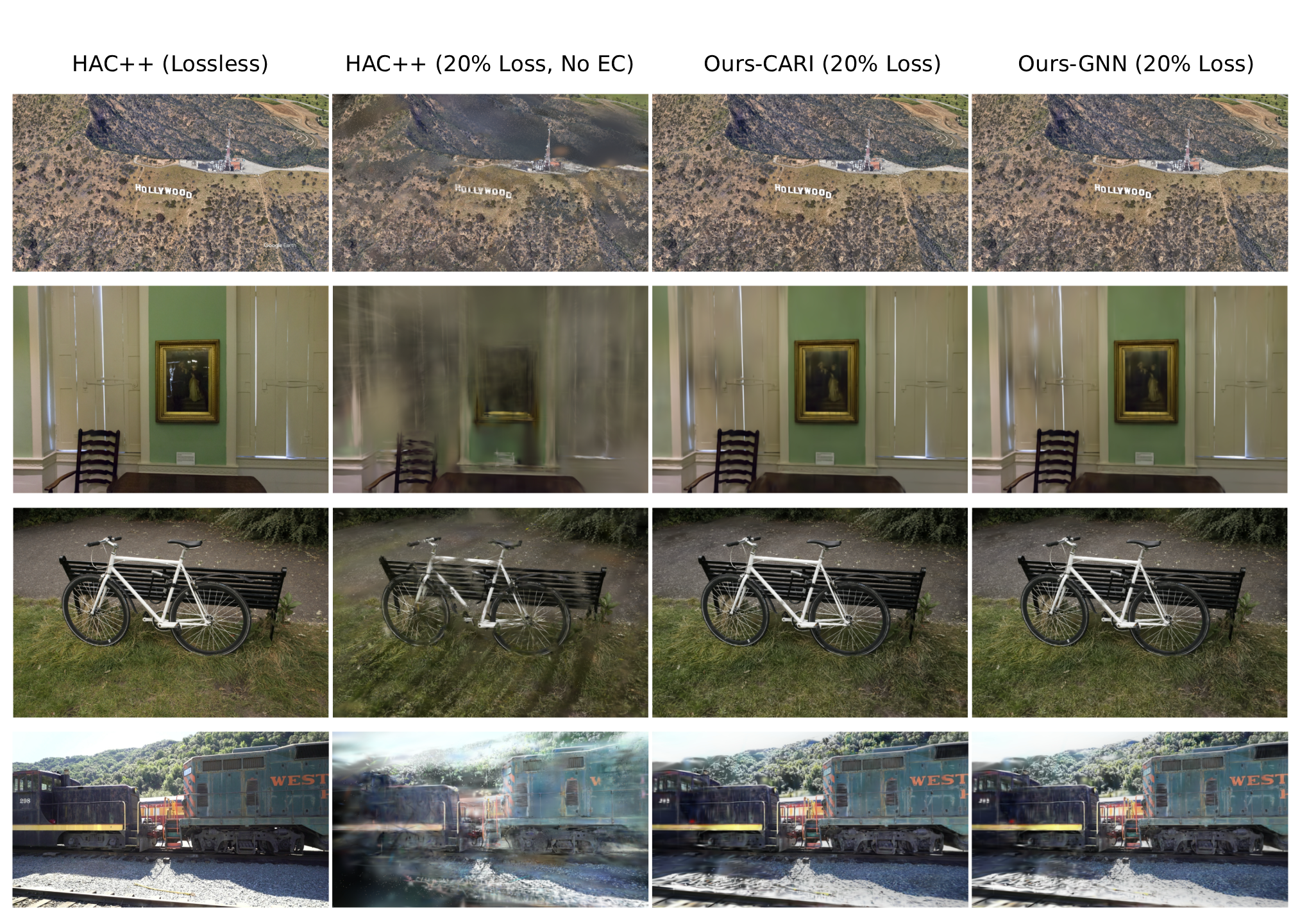}
    \Description{Qualitative comparison grid showing representative rendered views under lossless HAC++, HAC++ with packet loss and no concealment, the proposed CARI concealment, and the proposed GNN concealment.}
    \caption{Qualitative comparison under 20\% random packet loss across representative scenes. Columns follow the labels shown in the figure: HAC++ reference without packet loss, HAC++ under 20\% loss without error concealment, Ours-CARI, and Ours-GNN.}
    \label{fig:qualitative_results}
\end{figure*}

Qualitative visual comparisons are presented in Fig.~\ref{fig:qualitative_results}. Each row shows a representative scene, with columns labeled as \textit{HAC++ (Reference)}, \textit{HAC++ (20\% Loss, No EC)}, \textit{Ours-CARI (20\% Loss)}, and \textit{Ours-GNN (20\% Loss)}, where EC denotes error concealment. The direct-drop setting exhibits severe blur, ghosting, and structural discontinuities. The proposed CARI branch restores the global scene layout and suppresses most packet-loss artifacts, while the proposed GNN branch further improves local texture and edge fidelity in complex regions.

\paragraph{Vulnerability of Original HAC++}
As demonstrated in Table~\ref{tab:quantitative_results_wide}, the reconstruction quality of the original HAC++ suffers a dramatic decline under packet loss. At the 20\% loss rate, the native attribute-separated transmission mode leads to roughly 9 dB average PSNR degradation across datasets. The root cause of this systemic collapse lies in its attribute-independent packaging mechanism: when packet loss occurs, the renderer is forced to process ``fragmented anchors'' with mismatched attribute parameters, leading to severe geometric distortion and color artifacts.

\paragraph{Effectiveness and Limitations of the Proposed CARI Interpolation Branch}
The proposed \textit{Context-Aware Residual Interpolation (CARI)} method, operating atop our anchor-level atomic packaging and stratified random grouping strategies, significantly enhances the system's resilience. At 20\% loss, CARI reaches 23.71 dB on BungeeNeRF, 25.73 dB on Mip-NeRF 360, and 22.21 dB on Tanks \& Temples, validating the effectiveness of hash-context prior-guided interpolation. However, this heuristic branch still tends to over-smooth high-frequency details in complex real-world scenes, leaving room for learned residual refinement.

\paragraph{Complementarity of the GNN Branch}
For the 20\% packet-loss setting summarized in Table~\ref{tab:quantitative_results_wide}, the feature-focused GNN configuration maintains an average PSNR of 24.31 dB on BungeeNeRF and 26.13 dB on Mip-NeRF 360, corresponding to PSNR drops of 4.19 dB and 2.97 dB, respectively. Compared with the proposed interpolation/CARI branch under the same packetization protocol, the GNN recovers an additional 0.60 dB on BungeeNeRF and 0.40 dB on Mip-NeRF 360. On Tanks \& Temples, the GNN slightly improves PSNR but CARI remains better in SSIM and LPIPS, indicating that learned residuals are not uniformly superior for all scenes and all perceptual metrics. One likely reason is that the evaluated Tanks \& Temples scenes contain relatively simpler local geometry and fewer highly complex texture transitions than the large-scale BungeeNeRF scenes; in such smooth or low-frequency regions, residual interpolation can preserve structural consistency with fewer learned-regression artifacts, which benefits SSIM and LPIPS. This observation motivates our conservative confidence-controlled design: GNN residuals are used mainly to refine high-frequency feature details in complex scenes, while feature, scale, and offset attributes all fall back to interpolation when their confidence is insufficient. In the final reported setting, scale and offset are kept on the interpolation branch by default because geometry errors are visually more destructive, whereas high-confidence feature residuals can still be blended.

\subsection{Ablation Studies}

To rigorously evaluate the individual contributions of the core modules in our proposed error concealment framework---namely \textit{Anchor-level Atomic Packaging}, \textit{Stratified Random Grouping}, and \textit{GNN-based Repair}---we conducted a step-by-step ablation study under a 20\% random packet loss setting. Table~\ref{tab:ablation_results} reports dataset-wide averages for BungeeNeRF, Mip-NeRF 360, and Tanks \& Temples.

\begin{table*}[htbp]
\centering
\caption{Ablation results of key modules under 20\% random packet loss. Each row isolates one packetization or concealment component. All values are dataset-wide averages. ``Strat. Random'' denotes Stratified Random Grouping.}
\label{tab:ablation_results}
\resizebox{\textwidth}{!}{
\begin{tabular}{@{}clll|ccc|ccc|ccc@{}}
\toprule
 & & & & \multicolumn{3}{c|}{\textbf{BungeeNeRF}} & \multicolumn{3}{c|}{\textbf{Mip-NeRF 360}} & \multicolumn{3}{c}{\textbf{Tanks \& Temples}} \\ 
\textbf{No.} & \textbf{Packaging} & \textbf{Grouping} & \textbf{Inpainting} & \textbf{PSNR $\uparrow$} & \textbf{SSIM $\uparrow$} & \textbf{LPIPS $\downarrow$} & \textbf{PSNR $\uparrow$} & \textbf{SSIM $\uparrow$} & \textbf{LPIPS $\downarrow$} & \textbf{PSNR $\uparrow$} & \textbf{SSIM $\uparrow$} & \textbf{LPIPS $\downarrow$} \\ \midrule
 
(a) & Separated & Morton (Z-order) & None & 19.55 & 0.610 & 0.380 & 20.05 & 0.590 & 0.405 & 15.20 & 0.520 & 0.455 \\
(b) & Atomic & Morton (Z-order) & None & 21.60 & 0.735 & 0.305 & 21.40 & 0.700 & 0.355 & 20.30 & 0.665 & 0.315 \\
(c) & Atomic & Strat. Random & None & 22.05 & 0.765 & 0.285 & 24.05 & 0.770 & 0.300 & 20.65 & 0.730 & 0.285 \\
(d) & Atomic & Strat. Random & CARI & 23.71 & 0.818 & 0.220 & 25.73 & 0.807 & 0.264 & 22.21 & \textbf{0.800} & \textbf{0.233} \\
(e) & \textbf{Atomic} & \textbf{Strat. Random} & \textbf{GNN (Ours)} & \textbf{24.31} & \textbf{0.827} & \textbf{0.205} & \textbf{26.13} & \textbf{0.811} & \textbf{0.258} & \textbf{22.32} & 0.796 & 0.237 \\ \bottomrule
\end{tabular}
}
\end{table*}

\paragraph{Impact of Atomic Packaging}
Comparing experiments (a) and (b), replacing the native ``attribute-separated packaging'' of HAC++ with our ``atomic packaging'' strategy---without any repair algorithm and maintaining Morton spatial ordering---yields clear PSNR gains across the evaluated datasets. The original HAC++ failure mode loses roughly 9 dB on average because attribute packets are decoded inconsistently, whereas atomic packaging converts this failure into cleaner complete-anchor erasures. This result supports our core hypothesis: forcing the degradation to converge into an ``all-or-nothing'' complete anchor loss is essential to eliminate the ``fragmented anchors'' that cause parameter chaos under alpha blending.

\paragraph{Impact of Stratified Random Grouping without Inpainting}
Experiment (c) further replaces Morton ordering with the proposed stratified random grouping while still disabling decoder-side inpainting. Under independent random packet loss, this row should not be interpreted as the final benefit of stratified random grouping by itself; cleanly missing anchors remain visible when no recovery is applied. Its role is to create spatially dispersed support patterns that are more suitable for subsequent interpolation and GNN repair. Consistent with this interpretation, Atomic + Stratified Random + None remains approximately 1.5--2.0 dB below the CARI branch, leaving a clear gap for decoder-side concealment.

\paragraph{Evolution of Error Concealment Methods}
Finally, we evaluate different concealment mechanisms under the same encoder configuration (Atomic + Stratified Random). Introducing \textit{CARI} in experiment (d) provides a strong and stable interpolation baseline by aggregating neighboring residuals with hash-prior guidance. However, pure interpolation tends to over-smooth edges in complex topologies. The confidence-controlled GNN framework in experiment (e) further improves the average PSNR from 23.71 dB to 24.31 dB on BungeeNeRF and from 25.73 dB to 26.13 dB on Mip-NeRF 360, reducing the corresponding average PSNR drops to 4.19 dB and 2.97 dB, respectively. On Tanks \& Temples, GNN slightly improves PSNR from 22.21 dB to 22.32 dB but CARI remains better on SSIM and LPIPS, which is consistent with the smoother scene content where interpolation can be more reliable than learned residual refinement. This ``interpolation for baseline, GNN for feature refinement'' hybrid strategy is critical for preserving details while maintaining fallback stability across diverse scene complexities.

\subsubsection{Performance Analysis across Scene Scales}
During the evaluation of experiment (d) (CARI) and experiment (e) (GNN), we observed a critical phenomenon: the performance gain of the GNN module is highly correlated with scene scale and complexity. In large-scale, complex scenes (e.g., exceeding 300,000 anchors), the GNN can outperform CARI interpolation because local feature aggregation and hash-prior extraction help restore high-frequency details in regions dominated by complex topology.

Conversely, in smaller scenes or highly smooth local regions, spatial interpolation (CARI) approaches the optimal solution. In these low-frequency regions, forced non-linear regression via GNN may introduce slight regression noise due to over-parameterization, occasionally leading to metrics slightly lower than interpolation. 

These scale-dependent performance variations justify the necessity of our \textit{attribute-wise confidence control and fallback} mechanism (Section~\ref{subsec:gnn_inpainting}). By dynamically evaluating the confidence of each predicted attribute and blending with or falling back to the interpolation baseline when needed, our framework achieves complementary advantages: unleashing the GNN's potential for high-frequency details in complex areas while preserving the robust interpolation estimate in uncertain or extremely sparse regions. This scale-adaptive design ensures an optimal balance between visual quality and system stability across any scene complexity.

\section{Conclusion and Limitations}
\label{sec:conclusion}

\textbf{Conclusion.} In this paper, we presented a robust transmission and error concealment framework for 3D Gaussian Splatting (3DGS), systematically addressing the severe rendering degradation caused by packet loss in real-world network streaming. By reshaping the failure mode at the encoder side through \textit{Anchor-level Atomic Packaging} and \textit{Stratified Random Grouping}, we reduce catastrophic artifacts caused by attribute mismatches and avoid intractable large-scale spatial voids. At the decoder side, we proposed a prior-aware error concealment pipeline that integrates Context-Aware Residual Interpolation (CARI) with a lightweight two-layer GNN calibrated offline for each scene. Empowered by attribute-wise confidence control and fallback, our framework adaptively recovers high-frequency details while preserving rendering stability when individual attribute predictions are unreliable. Experiments under 20\% random packet loss demonstrate that our method substantially improves over no-concealment transmission and that the GNN branch provides complementary gains over CARI in complex regions while safely falling back to interpolation in low-confidence cases.

\textbf{Limitations and Future Work.} Despite the substantial improvements in transmission robustness, our current framework has certain limitations that warrant future research:
\begin{itemize}
    \item \textbf{Compression Efficiency Trade-off:} To achieve high error resilience, our atomic packaging and stratified random grouping strategies intentionally break the continuous Morton-order sorting and attribute-decoupled clustering utilized by the original HAC++ codec. In the current implementation, the all-non-weight bitstream size increases by 2.99\% across the evaluated scenes, while the attribute-only packet files increase by 7.13\%. Future work should further optimize this robustness--bitrate trade-off.
    \item \textbf{Reliability of Global Side Information:} The current evaluation assumes that global side information, including anchor coordinates, hash-grid/MLP parameters, and interleaving metadata, is delivered reliably through retransmission, forward error correction, or a protected control channel. A fully unreliable end-to-end channel would require joint protection of both global metadata and per-anchor attribute packets.
    \item \textbf{Computational Overhead at Decoder:} Although we explicitly designed a lightweight two-layer GNN module, the dynamic $k$-NN bipartite graph construction and network inference introduce additional computational latency. This could pose challenges for strict real-time rendering (e.g., >60 FPS) on resource-constrained edge devices or mobile platforms.
\end{itemize}

\textbf{Future Directions.} Future efforts will focus on exploring a joint Rate-Distortion-Resilience (RDR) optimization framework to strike an optimal balance between compression ratio and error concealability. Furthermore, we plan to investigate hardware-accelerated (e.g., highly optimized CUDA kernels) repair modules and the end-to-end co-training of the entropy codec alongside the error concealment network.
\bibliographystyle{ACM-Reference-Format}
\bibliography{reference}

@article{kerbl20233d,
  title={3d gaussian splatting for real-time radiance field rendering.},
  author={Kerbl, Bernhard and Kopanas, Georgios and Leimk{\"u}hler, Thomas and Drettakis, George and others},
  journal={ACM Trans. Graph.},
  volume={42},
  number={4},
  pages={139--1},
  year={2023}
}

@article{fan2024lightgaussian,
  title={Lightgaussian: Unbounded 3d gaussian compression with 15x reduction and 200+ fps},
  author={Fan, Zhiwen and Wang, Kevin and Wen, Kairun and Zhu, Zehao and Xu, Dejia and Wang, Zhangyang},
  journal={Advances in neural information processing systems},
  volume={37},
  pages={140138--140158},
  year={2024}
}

@inproceedings{niedermayr2024compressed,
  title={Compressed 3d gaussian splatting for accelerated novel view synthesis},
  author={Niedermayr, Simon and Stumpfegger, Josef and Westermann, R{\"u}diger},
  booktitle={Proceedings of the IEEE/CVF Conference on Computer Vision and Pattern Recognition},
  pages={10349--10358},
  year={2024}
}

@inproceedings{lu2024scaffold,
  title={Scaffold-gs: Structured 3d gaussians for view-adaptive rendering},
  author={Lu, Tao and Yu, Mulin and Xu, Linning and Xiangli, Yuanbo and Wang, Limin and Lin, Dahua and Dai, Bo},
  booktitle={Proceedings of the IEEE/CVF conference on computer vision and pattern recognition},
  pages={20654--20664},
  year={2024}
}

@inproceedings{chen2024hac,
  title={Hac: Hash-grid assisted context for 3d gaussian splatting compression},
  author={Chen, Yihang and Wu, Qianyi and Lin, Weiyao and Harandi, Mehrtash and Cai, Jianfei},
  booktitle={European Conference on Computer Vision},
  pages={422--438},
  year={2024},
  organization={Springer}
}

@article{chen2025hac,
  title={Hac++: Towards 100x compression of 3d gaussian splatting},
  author={Chen, Yihang and Wu, Qianyi and Lin, Weiyao and Harandi, Mehrtash and Cai, Jianfei},
  journal={IEEE Transactions on Pattern Analysis and Machine Intelligence},
  year={2025},
  publisher={IEEE}
}

@article{wang2000error,
  title={Error control and concealment for video communication: A review},
  author={Wang, Yao and Zhu, Qin-Fan},
  journal={Proceedings of the IEEE},
  volume={86},
  number={5},
  pages={974--997},
  year={1998},
  doi={10.1109/5.664283}
}

@article{wenger2003h,
  title={H.264/AVC over IP},
  author={Wenger, Stephan},
  journal={IEEE Transactions on Circuits and Systems for Video Technology},
  volume={13},
  number={7},
  pages={645--656},
  year={2003},
  doi={10.1109/TCSVT.2003.814966}
}

@article{goyal2001multiple,
  title={Multiple description coding: Compression meets the network},
  author={Goyal, Vivek K.},
  journal={IEEE Signal Processing Magazine},
  volume={18},
  number={5},
  pages={74--93},
  year={2001},
  doi={10.1109/79.952806}
}

@article{wang2019dynamic,
  title={Dynamic graph CNN for learning on point clouds},
  author={Wang, Yue and Sun, Yongbin and Liu, Ziwei and Sarma, Sanjay E. and Bronstein, Michael M. and Solomon, Justin M.},
  journal={ACM Transactions on Graphics},
  volume={38},
  number={5},
  year={2019}
}

@inproceedings{qian2021pu,
  title={Pu-gcn: Point cloud upsampling using graph convolutional networks},
  author={Qian, Guocheng and Abualshour, Abdulellah and Li, Guohao and Thabet, Ali and Ghanem, Bernard},
  booktitle={Proceedings of the IEEE/CVF conference on computer vision and pattern recognition},
  pages={11683--11692},
  year={2021}
}

@inproceedings{yu2021pointr,
  title={Pointr: Diverse point cloud completion with geometry-aware transformers},
  author={Yu, Xumin and Rao, Yongming and Wang, Ziyi and Liu, Zuyan and Lu, Jiwen and Zhou, Jie},
  booktitle={Proceedings of the IEEE/CVF international conference on computer vision},
  pages={12498--12507},
  year={2021}
}

@inproceedings{barron2022mip,
  title={Mip-nerf 360: Unbounded anti-aliased neural radiance fields},
  author={Barron, Jonathan T and Mildenhall, Ben and Verbin, Dor and Srinivasan, Pratul P and Hedman, Peter},
  booktitle={Proceedings of the IEEE/CVF conference on computer vision and pattern recognition},
  pages={5470--5479},
  year={2022}
}

@article{knapitsch2017tanks,
  title={Tanks and temples: Benchmarking large-scale scene reconstruction},
  author={Knapitsch, Arno and Park, Jaesik and Zhou, Qian-Yi and Koltun, Vladlen},
  journal={ACM Transactions on Graphics (ToG)},
  volume={36},
  number={4},
  pages={1--13},
  year={2017},
  publisher={ACM New York, NY, USA}
}

@article{hedman2018deep,
  title={Deep blending for free-viewpoint image-based rendering},
  author={Hedman, Peter and Philip, Julien and Price, True and Frahm, Jan-Michael and Drettakis, George and Brostow, Gabriel},
  journal={ACM Transactions on Graphics (ToG)},
  volume={37},
  number={6},
  pages={1--15},
  year={2018},
  publisher={ACM New York, NY, USA}
}

@inproceedings{xiangli2022bungeenerf,
  title={Bungeenerf: Progressive neural radiance field for extreme multi-scale scene rendering},
  author={Xiangli, Yuanbo and Xu, Linning and Pan, Xingang and Zhao, Nanxuan and Rao, Anyi and Theobalt, Christian and Dai, Bo and Lin, Dahua},
  booktitle={European conference on computer vision},
  pages={106--122},
  year={2022},
  organization={Springer}
}

@article{wang2004ssim,
  author={Wang, Zhou and Bovik, Alan C. and Sheikh, Hamid R. and Simoncelli, Eero P.},
  title={Image Quality Assessment: From Error Visibility to Structural Similarity},
  journal={IEEE Transactions on Image Processing},
  year={2004},
  volume={13},
  number={4},
  pages={600--612},
  month={Apr.},
  doi={10.1109/TIP.2003.819861}
}

@inproceedings{zhang2018lpips,
  author={Zhang, Richard and Isola, Phillip and Efros, Alexei A. and Shechtman, Eli and Wang, Oliver},
  title={The Unreasonable Effectiveness of Deep Features as a Perceptual Metric},
  booktitle={Proceedings of the IEEE/CVF Conference on Computer Vision and Pattern Recognition (CVPR)},
  year={2018},
  pages={586--595},
  doi={10.1109/CVPR.2018.00068}
}

\end{document}